\documentclass[twocolumn]{revtex4}
\usepackage{color}
\usepackage{colordvi}
\usepackage{amsmath}
\usepackage{amssymb}
\usepackage{epsfig}
\newcommand{\rv}{\mathbf{r}}

\newcommand{\Rc}{\mathcal{R}}

\begin{document}

\title{Accurate, efficient and simple forces with Quantum Monte Carlo methods}

\author{Simone Chiesa}\email{chiesa@uiuc.edu}
\affiliation{ Dept. of Physics, University of Illinois
Urbana-Champaign, Urbana, IL 61801}
\author{D. M. Ceperley}\email{ceperley@uiuc.edu}
\affiliation{ Dept. of Physics, University of Illinois
Urbana-Champaign, Urbana, IL 61801}
\author{Shiwei Zhang}\email{shiwei@physics.wm.edu}
\affiliation{ Dept. of Physics, College of William and Mary,
Williamsburg, VA 23187}

\begin{abstract}
Computation of ionic forces using quantum Monte Carlo methods has long been
a challenge.
We introduce a simple procedure,
based on known properties of physical
electronic densities, to make the variance of the Hellmann-Feynman
estimator finite.
We obtain very accurate geometries for the molecules H$_2$, LiH, CH$_4$,
NH$_3$, H$_2$O and HF, with a Slater-Jastrow trial wave function.
Harmonic frequencies for diatomics are also in good agreement with
experiment.
An antithetical sampling method is also discussed for additional 
reduction of the variance.
\end{abstract}
\maketitle

The optimization of molecular geometries and crystal structures
and {\em ab initio} molecular dynamics simulations are among the most
significant achievements of single particle theories. These
accomplishments were both possible thanks to the possibility of
readily computing forces on the ions within the framework of the
Born-Oppenheimer approximation.
The approximate treatment of electron interactions typical of these
approaches can, however,
lead to quantitatively, and sometimes qualitatively, wrong
results. This fact, together with a favorable scaling
of the computational cost with respect to the number of 
particles, has
spurred the development of  stochastic techniques, {\em i.e.}
quantum Monte Carlo (QMC) methods. Despite the higher accuracy
achievable for many physical properties, the lack of an efficient
estimator for forces has prevented, until
recently\cite{Filippi,Casalegno,Assaraf2}, the use of QMC
methods to predict even the simplest molecular geometry. 
The chief problem is to have a Monte Carlo (MC) estimator for the force
with sufficiently small variance. For example, in
all-electron calculations, a
straightforward application of MC sampling of the
Hellmann-Feynman estimator has infinite variance.
This can be easily seen from the definition of the force.
For a nucleus of charge $Z$ at the origin, the force
can be written, together with its variance, as a function of the charge 
density $\rho(\rv)$ as
\begin{equation}
\mathbf{F}= Z \int d\rv\rho(\rv)\frac{\rv}{r^3}\:;\qquad
\sigma^2= Z^2 \int d\rv\rho(\rv)\frac{1}{r^4}-F^2 .
\label{HF_def}
\end{equation}
Since the electronic density is finite
at the origin, the variance integral diverges. 

\begin{figure}
\vskip 0.5cm
\begin{center}
\epsfig{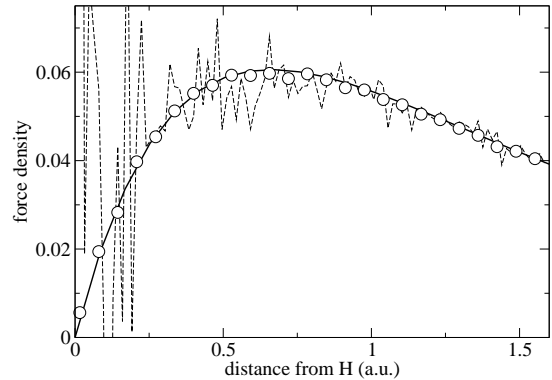}
\caption{Force density along the z-direction for the H atom in LiH. The 
bond is along the z-axis, with a length of 3.316 Bohr. 
The continuous black curve is calculated
from the Hartree-Fock orbitals. 
The dashed line is the estimate of $f_z$ using the bare estimator. 
Circles
are obtained in an identical QMC simulation using the antithetic sampling
technique outlined in the text.}
\label{forcedens}
\end{center}
\end{figure}

In this paper, we propose a modified form for the force estimator
which has finite variance.
This estimator is then used to calculate forces and predict
equilibrium geometry and vibrational frequencies for a set of
small molecules.
Without loss of generality we will consider only the $z$-component
of the  force on an atom at the origin. 
In a QMC calculation based in configuration space, the charge density is
a sum of delta functions: $\rho(\rv)\propto \sum_{\rv'} \delta(\rv-\rv')$,
where the sum is over all $N_e$ electron positions and all MC samples.
We consider separately the electrons within a distance $\Rc$ of the atom
and those outside. 
The contribution to the force from charges outside, $F_z^O$,
can be calculated directly 
with the Hellmann-Feynman estimator in Eq.~(\ref{HF_def}).
The contribution from inside the sphere is responsible for the 
large variances in the direct estimator.
It is convenient to introduce a ``force
density'' defined as the force arising from electron charges at a distance
$r$ from the origin:
\begin{equation}
f_z(r)=Z\int d\Omega\; \rho(r,\theta,\phi) \cos\theta
\label{frcd_def}
\end{equation}
Then the force is given as:
 \begin{equation} \label{fsum}
F_z=F_z^O+ \int_0^{\Rc} f_z(r)dr.
\end{equation}
The force density is a smooth function of $r$ that tends to $0$
linearly as $r$ approaches the origin. The force density for H in
a LiH molecule computed with Hartree-Fock and two different QMC estimators
is shown in Fig.\ref{forcedens}.  As expected the bare force
estimator fluctuates wildly at small $r$.

Because the force density is a smooth function, we can represent
it in the interval $(0,\Rc)$ with a polynomial
\begin{equation}
\widetilde{f_z}(r)=\sum_{k=1}^M a_k r^k \label{polynom}
\end{equation}
and determine the coefficients, $a_k$, by minimizing
\begin{equation}
\chi^2=\int_0^{\Rc} dr\,r^m\left[ f_z(r) - \widetilde{f_z}(r) \right]^2
\label{chi2}
\end{equation}
where $r^m$ is a weight factor used to balance contributions from
different values of $r$.
\begin{figure}[b]
\vskip 0.5cm
\begin{center}
\epsfig{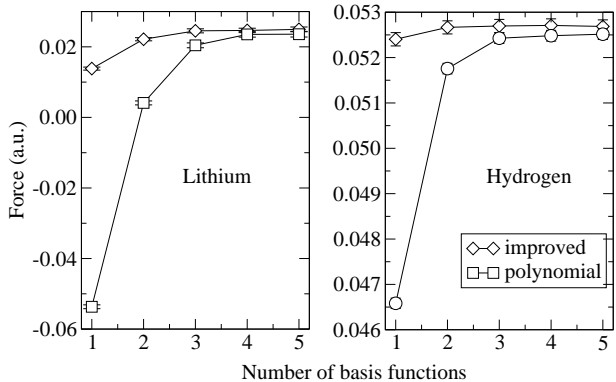}
\caption{Dependence of the VMC force on the expansion basis, 
for LiH with a bond length of 3.316 Bohr. The fitting radius $\Rc=$0.6
Bohr. The definitions of the basis functions are in
Eq.'s~(\ref{polynom}) and (\ref{modpolyn}). The
forces on H and Li are different because of the lack of full
optimization of the VMC wave function (see text).} \label{LiH_fit}
\end{center}
\end{figure}

Since the relation between the force and the force density is
linear, and the relation between the fitting coefficients and the
electronic density is linear, we can directly write the force as
averages over moments of the force density. After some
manipulations we arrive at:
\begin{equation}
F_z=F_z^O+ Z\left\langle \sum_{i=1}^{N_e} g(r_i)\frac{z_i}{r_i^3}\right\rangle_{\rm MC},
\label{finalforce}
\end{equation}
where the new estimator function is:
 \begin{equation}
 g(r)=\theta(\Rc -r)\sum_{k=1}^M c_k r^{k+m}.
 \end{equation}
The coefficients  $c_k$'s are determined by
$\mathbf{c}=\mathbf{S^{-1}h}$ where the Hilbert matrix $\mathbf{S}$  and
the residual vector $\mathbf{h}$ are
\begin{equation}
S_{kj}=\frac{\Rc^{m+k+j+1}}{m+k+j+1},\qquad
h_j=\frac{\Rc^{j+1}}{j+1}. \label{Sh}
\end{equation}
Note that for the bare estimator $g(r)=\theta(\Rc -r)$. Because of the
restriction on the basis, the variance of the new estimator is finite as 
long as $m>-1/2$. We have numerically found that the weighting factor $m=2$,
where each volume element is weighted equally, gives the lowest
variance estimate of the force.

\begin{figure}
\begin{center}
\epsfig{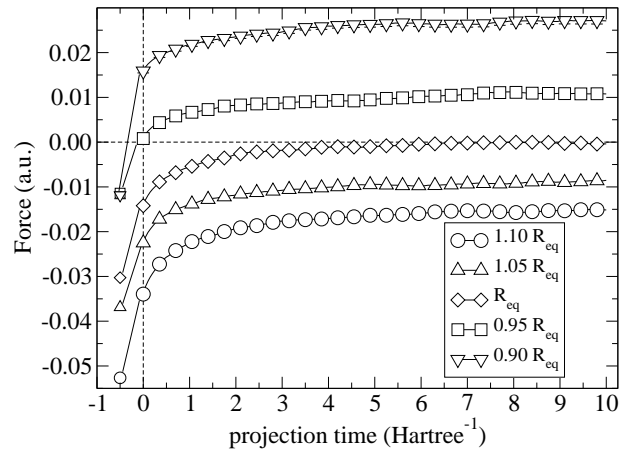}
\caption{Projection of the force in LiH using forward walking. The points
at negative imaginary time give the VMC values. Values at $0$ are the
mixed estimates of the DMC simulation.}
\label{LiH_fw}
\end{center}
\end{figure}

To derive the estimator we have used the fact that $f_z(r)$ goes linearly at small $r$.
\footnote{The force density $f_z(r)$ is proportional to the $p_z$ component of
the density. A non-zero value as $r\rightarrow 0$ would
imply a discontinuity of $\rho$ at the origin 
along the $z$-direction.} 
This is the crucial property that allows to {\em filter} out the $s$-wave
component of the density responsible for the variance divergence.
The original estimator is correct for any arbitrary
charge density while the new filtered one uses physical
properties of the charge density to reduce the variance.
The variance depends on the fitting radius $\Rc$ and on the basis
set size $M$. As $\Rc$ increases, the size of the basis must increase,
which increases the variance.  
Charge densities corresponding to low energy states must be smooth
and we typically find that only 2 or 3
basis functions are needed. The size of the basis can be reduced
by using more appropriate basis sets. For example, in all
calculations reported below we used the expansion
\begin{equation}
\widetilde{f}_z(r)=f^\text{SD}_z(r)\sum_{k=0}^{M} a_k r^k, 
\label{modpolyn}
\end{equation}
where $f^\text{SD}_z$ is the force density of a single determinant wave
function, which can be readily computed from the orbitals. The
improved basis allows a smaller polynomial set and a reduction of the
variance. In Fig.~\ref{LiH_fit} the dependence of the bias on the
basis set type and size is shown for
the case of a variational Monte Carlo (VMC) simulation on LiH at a bond length of $3.316$ Bohr.

The trial wave functions $\Psi_T$ used in all cases were of the
Slater-Jastrow form. The orbitals were obtained from a
Hartree-Fock calculation using CRYSTAL98 \cite{Crys98}. The
electron-electron and electron-proton Jastrow factors had the form of
$\exp(ar/(1+br))$, with $a$ and $b$ optimized by minimizing
$|E_{loc}-\langle E \rangle|$ \cite{Bressa} over
points sampled from $|\Psi_T|^2$. The time step in the 
diffusion Monte Carlo (DMC)
simulations was chosen to give an acceptance ratio of $98$\%, a
value for which the time step bias on forces was within the statistical error
bars.

\begin{table}
\caption{Equilibrium distances in \AA. 
Experimental, CCSD(T) and B3LYP values were taken from Ref\cite{CCCBDB}.
The CCSD(T) and the B3LYP 
results were obtained using the cc-pVTZ basis set with the exception of LiH 
where the 6-311G* set was used. PBE results \cite{Xu_Goddard} were all obtained
using the aug-cc-pVTZ basis set.}

\begin{ruledtabular}
\begin{tabular}{lccccc}
                & QMC       & Exp.  &CCSD(T)& B3LYP    & PBE \\
\hline
H$_2$           & 0.7419(4) & 0.741 & 0.743 & 0.743    & 0.751\\
LiH             & 1.592(4)  & 1.596 & 1.618 & 1.595    & 1.606\\
CH$_4$          & 1.091(1)  & 1.094 & 1.089 & 1.088    & 1.096\\
NH$_3$ (N-H)    & 1.009(2)  & 1.012 & 1.014 & 1.014    & 1.023\\
NH$_3$ (H-H)    & 1.624(2)  & 1.624 & 1.616 & 1.624    & 1.634\\
H$_2$O (O-H)    & 0.959(2)  & 0.956 & 0.959 & 0.961    & 0.971\\
H$_2$O (H-H)    & 1.519(3)  & 1.517 & 1.508 & 1.520    & 1.531\\
HF              & 0.919(1)  & 0.918 & 0.917 & 0.923    & 0.932
\end{tabular}
\end{ruledtabular}
\label{geomtab}
\end{table}

Since the exact density is needed for the Hellmann-Feynman theorem, forward
walking\cite{Hammond_book} or one of the variational path integral
algorithms\cite{PI_Cep,Baroni} is needed in order to evaluate the
force estimator. An example of the convergence of forward walking
is shown in Fig.~\ref{LiH_fw}. The force as a function of the
forward-walking projection time quickly reaches a plateau
corresponding to the exact value. In this example, the variational 
forces are far from correct. This discrepancy
results from the lack of full optimization of the trial wave function
made of localized basis orbitals and atom centered Jastrow factors,
and can be reduced somewhat by including Pulay terms \cite{Casalegno}.
In DMC, forward walking eliminates the need for the Pulay corrections.

\begin{figure}
\begin{center}
\epsfig{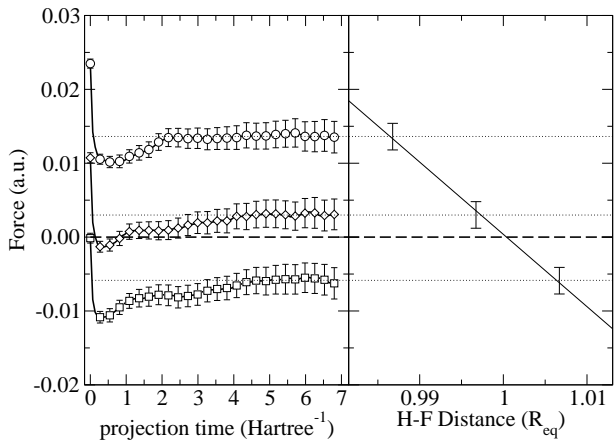}
\caption{DMC force in hydrogen fluoride. 
Left panel: evolution of the force over forward-walking time. Right panel:
fully projected forces as a function of nuclear distance. R$_\text{eq}$ is
the experimental equilibrium distance.}
\label{HF}
\end{center}
\end{figure}

The equilibrium geometries were computed by fitting the QMC forces in
the proximity of the equilibrium geometry to a
polynomial with the appropriate symmetry.
Fig.~\ref{HF} shows the force in hydrogen fluoride in a 2\%
interval around the equilibrium geometry. The equilibrium
geometries are reported in Table \ref{geomtab} together with those
given by CCSD(T), DFT using the B3LYP and the PBE functional, and experiments.
The differences between QMC and experimental values
are in all cases less than $0.4\%$ and closer to the
experiment than the other techniques. For diatomics it is easy
to provide an estimate of the harmonic vibrational frequencies
starting from the derivative of the force curve at equilibrium geometry.
The QMC frequencies, reported in Table \ref{freqtab}, are in good
agreement with the experiment, with errors comparable to that from
CCSD(T) and DFT PBE or B3LYP. This suggests that forces computed within our
approach are accurate also away from the equilibrium and
could be used in molecular dynamics calculations or to optimize
molecular geometries.

The only source of systematic error in our calculations that
cannot be simply addressed is the fixed-node error. In fixed-node 
DMC, the random walk is
forbidden to cross the nodes of the trial wavefunction in order to prevent the
loss of efficiency due to the fermion antisymmetry.
If the nodes are accurate, so is the QMC energy and electronic
density; hence the force. For incorrect nodes, the energy is an
upper bound to the true energy, but such can not be said 
for the force. It is also not necessarily the case that the
forces obtained from Eq.~(1) are equal to the gradient of the
fixed-node energy\cite{Schautz1,Schautz2,Huang}: this is only
guaranteed in the limit of exact nodal surfaces. The high quality
of the geometries and vibrational frequencies suggests that these
errors, at least for the cases treated in this paper, are
negligible. This is perhaps not surprising, since the electronic
density is a 1-electron property, while the nodal error is a
many-body effect. 

\begin{table}
\caption{Harmonic frequencies in cm$^{-1}$. 
Experimental, CCSD(T) and B3LYP values were taken from Ref\cite{CCCBDB}.
The CCSD(T) and the B3LYP results were obtained
using the cc-pVTZ basis set with the exception of LiH where the 6-311G* set was used.
PBE results \cite{PPP} were obtained using {\em ad hoc} gaussian basis sets.}
\begin{ruledtabular}
\begin{tabular}{lccccc}
 & QMC & Exp. & CCSD(T) & B3LYP & PBE \\
\hline
H$_2$           & 4464(18)  & 4410  & 4420 & 4401 & 4323 \\
LiH             & 1445(20)  & 1369  & 1414 & 1405 & 1380 \\
HF              & 4032(266) & 4181  & 4085 & 4138 & 4001
\end{tabular}
\end{ruledtabular}
\label{freqtab}
\end{table}

We have also tested another method to further reduce the variance of the
Hellmann-Feynman estimator.
The filtered estimator performs well on the hydrogen atom but
for heavier nuclei the error bar grows and seems to scale as $Z^3$.
In those cases the new method can potentially be very useful, with
error bars scaling between $Z$ and $Z^2$. 
The method is based on the observation that,
while electrons in the core cause large fluctuations in the force
density, they contribute very little to it. A standard approach to
reduce the variance of a Monte Carlo estimate is the use of
antithetic variates\cite{kaloswhitlock}: a positive fluctuation is
paired with a negative fluctuation. 
Suppose the random walk arrives at
a multidimensional electronic configuration $R$, 
with $p$  ($\ge 1$) electrons inside
a radius $\Rc_{av}\le\Rc$ of an atom located at the origin. 
We obtain an antithetic
configuration $R'$ by reflecting all $p$ core electrons about
the origin.
We then estimate the force contribution due to the $p$ electrons using
both $R$ and $R'$, assigning a weight factor of 
$w(R')=|\psi(R')/\psi(R)|^2$ to $R'$.
Their joint contribution to the estimator in Eq.~(\ref{finalforce}) is
$Z\frac{1-w(R')}{2}\sum_i g(r_i) z_i r^{-3}$ where the sum runs over 
the $p$ core electrons. Since $w\rightarrow 1$ as $\Rc_{av}\rightarrow 0$, 
fluctuations in the core are much reduced.

Within VMC this scheme can be implemented
exactly, leading to a dramatic reduction of the variance
as can be noticed from
Fig.~\ref{forcedens}.
However
this estimator is non-local and, in DMC, suffers from the same
problems as non-local pseudopotentials, making an unbiased 
implementation not straightforward. 
We
postpone further discussion of the antithetic method to a
future article.

Two other approaches have been
introduced recently for the computation of forces in QMC. Filippi and
Umrigar have computed forces for diatomics by correlating random
walks for interatomic separations $a$ and $a'$. In DMC
the difficulty associated with the nodal error and the branching
factor was overcome by neglecting some types of correlation. The
main drawback of a finite difference method is the difficulty of
calculating all the components of the force simultaneously; for a system
of $N$ atoms this
method would require $3N$ separate force calculations.

The other approach, introduced in
Ref.~\cite{Assaraf2}, is closer to our method.
It is based on a ``zero-variance'' version of
the Hellmann-Feynman estimator and can be understood
in the framework of this paper: one can prove that 
it corresponds to filtering out the $s$-wave component 
of the density leaving the force density unchanged.
The semi-local character
of the ``zero-variance'' estimator makes its DMC implementation 
trickier. To overcome this problem
there have been attempts\cite{Casalegno,Assaraf3} to 
use correction terms similar in nature to the 
Pulay terms in single-particle approaches. In practice, this
scheme requires extensive optimization and, although promising, it is
unclear if it will be viable for more complicated
cases. In addition, the value of the force is very sensitive to
small errors\cite{Assaraf3} in the charge density and
the optimization within a stochastic technique is probably not
sufficiently stable to eliminate these errors.

In conclusion, we have developed a simple method for computing
forces within quantum Monte Carlo and used it to find
the equilibrium geometries for small polyatomic molecules.  This
has been the first time that a QMC technique is used to predict
geometries of molecules beyond diatomics. The only overhead in the
calculation is the necessity of determining unbiased estimators,
which requires the use of either forward-walking or reptation MC
techniques. The new 
method leads to very accurate forces 
despite errors from the fixed-node approximation and from its contribution to 
the energy derivatives.
Extension
of the method, including the antithetic estimator technique, 
to heavier atoms and to atoms
with pseudopotentials\footnote{The filtered estimator can be applied 
to atoms with non-local pseudopotentials, where the density is replaced 
by a density matrix and a corresponding force density can be defined by summing over 
partial waves.}is under investigation.

This material is based upon work supported in part by the U.S Army Research
Office under DAAD19-02-1-0176. Computational support was provided by the
Materials Computational Center and the National Center for Supercomputing
Applications at the University of Illinois. S.Z. acknowledges support from NSF.

\end{document}